\documentclass[onecolumn,secnumarabic,amssymb, nobibnotes, aps, prd]{revtex4}
\usepackage{bm}

\begin{document}

\title{Charmonium Mass Spectrum with Spin-Dependent Interaction\\ in Momentum-Helicity Space}%

\author{M. Radin}
\affiliation{Department of Physics, K. N. Toosi University of Technology, P.O.Box 16315--1618, Tehran, Iran.}%

\date{July 22, 2014}%

\begin{abstract}
In this paper we have solved the nonrelativistic form of the
Lippmann-Schwinger equation in the momentum-helicity space by
inserting a spin-dependent quark-antiquark potential model
numerically. To this end, we have used the momentum-helicity basis
states for describing a nonrelativistic reduction of one gluon
exchange potential. Then we have calculated the mass spectrum of the
charmonium $\psi(c\bar{c})$, and finally we have compared the
results with the other theoretical results and experimental data.
\end{abstract}


\maketitle

\section{Introduction}

During the past years, several models and methodological approaches
based on solving the relativistic and nonrelativistic form of the
Schr$\ddot{\mathrm{o}}$dinger or Lippmann-Schwinger~ equation have
been developed for studying the light and heavy mesons in the
coordinate and momentum spaces respectively.

Recently, the three-dimensional approach based on momentum-helicity
basis states for studding the Nucleon-Nucleon scattering and
deuteron state has been developed~\cite{h1,h2}. We extend this
approach to particle physics problems by solving the nonrelativistic
form of the Lippmann-Schwinger equation to obtain the mass spectrum
of the heavy messons using the nonrelativistic quark-antiquark
interaction in terms of a linear confinement, a Coulomb, and various
spin-dependent pieces.

In the heavy-quark (c,b) mesons the differences between energy
levels are small compared to the particle masses. Hence, the
nonrelativistic Lippmann-Schwinger equation can be used to study
their quantum behavior. To this end, we have used the
nonrelativistic form of the Lippmann-Schwinger equation in the
momentum-helicity representation to study the charmonium as a heavy
meson. For this purpose, we have used a nonrelativistic
quark-antiquark potential based on one-gluon exchange in the
momentum-helicity representation.

This article is organized as follows. In Sect.~2, the
nonrelativistic Lippmann–Schwinger equation in the momentum-helicity
basis states which leads to coupled and uncoupled integral equations
for various quantum numbers is presented briefly. In Sect.~3, a spin
dependent quark-antiquark potential model is described in the
momentum-helicity basis states. The details of the numerical
calculations and the results obtained for the charmonium are
presented in Sect.~4. Finally, a summary and an outlook are provided
in Sect.~5.

\section{Lippmann-Schwinger Equation in Momentum-Helicity basis states}
The nonrelativistic form of the homogenous Lippmann-Schwinger
equation for describing the heavy meson bound state is given by:
\begin{eqnarray}\label{eq6}
|\Phi_{j}^{M_{j}}\rangle=\frac{1}{E-\frac{p^{2}}{m}}\,V\,|\Phi_{j}^{M_{j}}\rangle,
\end{eqnarray}
where $V$ denotes the quark-antiquark interaction, $m$ is mass of
the quark or antiquark and $|\Phi_{j}^{M_{j}}\rangle$ is the meson
bound state with the total angular momentum $j$. $M_j$ is projection
of the total angular momentum $j$ along the quantization axis. The
integral form of this equation in the momentum-helicity basis states
is written as~\cite{RN}:
\begin{eqnarray}\label{eq8}
\mathrm{\Phi}_{Sj}^{M_{j}}(p)&=&\frac{2\pi}{E-\frac{p^{2}}{m}}\sum_{\mathrm{\Lambda}'}\int_{0}^{\infty}
dp'p'^{2}\,V^{S}_{M_{j}\mathrm{\Lambda}'}(p,p')\,\mathrm{\Phi}_{Sj}^{\mathrm{\Lambda}'}(p'),
\end{eqnarray}
with:
\begin{eqnarray}\label{eq8}
V^{S}_{M_{j}\mathrm{\Lambda}'}(p,p')=\int_{-1}^{1}d\cos
\theta'\,V^{S}_{M_{j}\mathrm{\Lambda}'}(p,p',\theta')\,d^{j}_{M_{j}\mathrm{\Lambda}'}(\theta'),\\\nonumber
\end{eqnarray}
where $p$ is the magnitude of the relative momentum of the quark and
antiquark, $S$ is the total spin of meson, $\Lambda$ is the spin
projection along the relative momentum and
$d^{j}_{M_{j}\mathrm{\Lambda}'}(\theta')$ are the rotation matrices.
For an arbitrary total angular momentum $j$, and singlet case of the
total spin state, Eq.~(2) leads to one equation:
\begin{eqnarray}\label{eq8}
\mathrm{\Phi}_{0j}^{M_{j}}(p)&=&\frac{2\pi}{E-\frac{p^{2}}{m}}\int_{0}^{\infty}
dp'p'^{2}\,V^{0}_{M_{j}0}(p,p')\mathrm{\Phi}_{0j}^{0}(p').
\end{eqnarray}
Also for $j=0$ and triplet case of the total spin state, Eq.~(2)
leads to one equation as:
\begin{eqnarray}\label{eq8}
\mathrm{\Phi}_{1j}^{0}(p)&=&\frac{2\pi}{E-\frac{p^{2}}{m}}\int_{0}^{\infty}
dp'p'^{2}\,V^{1}_{00}(p,p')\mathrm{\Phi}_{1j}^{0}(p').
\end{eqnarray}
For $S=1$ and $j>0$ it is more complicated. For example for $j=1$,
Eq.~(2) leads to one equation for channel $P$ and two coupled
equations for channels $S$ and $D$ as follows:
\begin{eqnarray}\label{eq8}
\mathrm{\Psi}_{111}(p)&=&\frac{2\pi}{E-\frac{p^{2}}{m}}\int_{0}^{\infty}
dp'p'^{2}\,\Big[V^{1}_{11}(p,p')-V^{1}_{-11}(p,p')\Big]\,\mathrm{\Psi}_{111}(p'),\,\,\,\,\,\,\quad\quad\quad\quad\qquad\qquad\qquad\qquad\qquad\qquad\quad\,\,
\end{eqnarray}
\begin{eqnarray}\label{eq8}
\mathrm{\Psi}_{011}(p)&=&\frac{2\pi}{E-\frac{p^{2}}{m}}\frac{1}{3}\int_{0}^{\infty}
dp'p'^{2}\Biggl\{\Big[2V^{1}_{11}(p,p')+2V^{1}_{01}(p,p')+V^{1}_{00}(p,p')+2V^{1}_{10}(p,p')+2V^{1}_{-11}(p,p')\Big]\,\mathrm{\Psi}_{011}(p')\nonumber\\&&+
\sqrt{2}\,\Big[V^{1}_{11}(p,p')+V^{1}_{01}(p,p')-V^{1}_{00}(p,p')-2V^{1}_{10}(p,p')+V^{1}_{-11}(p,p')\Big]\Biggl\}\mathrm{\Psi}_{211}(p'),
\end{eqnarray}
\begin{eqnarray}\label{eq8}
\mathrm{\Psi}_{211}(p)&=&\frac{2\pi}{E-\frac{p^{2}}{m}}\frac{1}{3}\int_{0}^{\infty}
dp'p'^{2}\Biggl\{\sqrt{2}\,\Big[V^{1}_{11}(p,p')-2V^{1}_{01}(p,p')-V^{1}_{00}(p,p')+V^{1}_{10}(p,p')+V^{1}_{-11}(p,p')\Big]\,\mathrm{\Psi}_{011}(p')\nonumber\\&&+
\Big[V^{1}_{11}(p,p')-2V^{1}_{01}(p,p')+2V^{1}_{00}(p,p')-2V^{1}_{10}(p,p')+V^{1}_{-11}(p,p')\Big]\Biggl\}\mathrm{\Psi}_{211}(p'),
\end{eqnarray}
where $\mathrm{\Psi}_{lSj}(p)$ is the partial wave component of the
wave function which is connected to the momentum-helicity component
of the wave function as~\cite{RN}:
\begin{eqnarray}\label{eq9}
\mathrm{\Phi}^{\mathrm{\Lambda}}_{jS}(p)&=&\sum_{l}\sqrt{\frac{2l+1}{4\pi}}\,C(lsj;0\mathrm{\Lambda}\mathrm{\Lambda})\,\mathrm{\Psi}_{lSj}(p).
\end{eqnarray}
The inverse relation is written as:
\begin{eqnarray}\label{eq9}
\mathrm{\Psi}_{lSj}(p)=\frac{\sqrt{4\pi
(2l+1)}}{2j+1}\sum_{\Lambda}\,C(lsj;0\mathrm{\Lambda}\mathrm{\Lambda})\,\mathrm{\Phi}^{\mathrm{\Lambda}}_{jS}(p).
\end{eqnarray}

\section{Quark-Antiquark Potential in Momentum-Helicity basis states}
The spin dependent potential model that we have used in our
calculations is sum of the Linear and a simple nonrelativistic
reduction of an effective one gluon exchange potential without
retardation. This potential in the coordinate space is given in
terms of~\cite{Sur}:
\begin{eqnarray}\label{eq8}
V(\textbf{r},\textbf{p})&=&\sigma
r+f_{c}\,\alpha_{s}\Bigg\{\frac{1}{r}-\frac{\pi}{m^2}\,\delta(\textbf{r})+\frac{1}{m^2}\frac{\textbf{p}\cdot\textbf{p}}{r}-\frac{3}{4m^{2}}\frac{\textbf{L}\cdot(\boldsymbol\sigma_{1}+\boldsymbol\sigma_{2})}{r^{3}}\nonumber\\&&-
\frac{2\pi}{3m^{2}}\,\delta(\textbf{r})(\boldsymbol\sigma_{1}\cdot\boldsymbol\sigma_{2})-
\frac{1}{4m^{2}}\frac{3(\boldsymbol\sigma_{1}\cdot\hat{\textbf{r}})(\boldsymbol\sigma_{1}\cdot\hat{\textbf{r}})-(\boldsymbol\sigma_{1}\cdot\boldsymbol\sigma_{2})}{r^{3}}\Bigg\},\quad\quad\quad
\end{eqnarray}
where $\sigma$ is the string tension, $\alpha_{s}$ is the
strong-interaction fine-structure constant, $f_{c}$ is the color
factor which is -4/3 for quark-antiquark and -2/3 for quark-quark,
$\boldsymbol\sigma_{1}$ and $\boldsymbol\sigma_{2}$ are the Pauli
matrices and $\textbf{L}$ is the total orbital angular momentum
operator. Fourier transformation of this potential to momentum space
yields:
\begin{eqnarray}\label{eq8}
\langle \textbf{p}| V|\textbf{p}'\rangle&=&\sigma
\,\Big[\,\delta(\textbf{q})\,r_{c}+\frac{1}{2\pi^{2}q^{4}}\big(2\cos(q\,r_{c})-2+q\,r_{c}\sin(q\,r_{c})\big)\Big]
\nonumber\\  &&+f_{c}\,\alpha_{s}\,e^{-\lambda^{2}
q^{2}}\Bigg\{\Big[\frac{\delta(\textbf{q})}{r_{c}}+\frac{1}{2\pi^{2}q^{2}}\big(1-\frac{\sin(q\,r_{c})}{q\,r_{c}}\big)\Big]\big(1+\frac{p^2}{m^{2}}\big)-\frac{1}{8\pi^{2}m^{2}}
\nonumber\\
&&+\frac{3}{8\pi^{2}m^{2}q^{2}}i(\boldsymbol\sigma_{1}+\boldsymbol\sigma_{2})\cdot\mathbf{p}\times
\textbf{p}'
 -\frac{1}{12\pi^{2}m^{2}}(\boldsymbol\sigma_{1}\cdot\boldsymbol\sigma_{2})+\frac{1}{24\pi^{2}m^{2}}\Big[3(\boldsymbol\sigma_{1}\cdot\hat{\textbf{q}})(\boldsymbol\sigma_{2}\cdot\hat{\textbf{q}})-(\boldsymbol\sigma_{1}\cdot\boldsymbol\sigma_{2})\Big]\Bigg\},
\end{eqnarray}
where $\textbf{q}=\textbf{p}'-\textbf{p}$ is the momentum transfer.
The kernels of integral equations have singularity. To overcome this
problem we have used the regularized form of linear confining and
Coulomb parts of the potential~\cite{ML}. Details of Fourier
transformation of regularized parts of the potential are given in
Appendix A. Also we have used a Gaussian form factor, $\exp
(-\frac{1}{2}\lambda^{2} q^{2})$ at the quark-gluon vertex as in
Ref.~\cite{J} to remove singularity of the kernels due to existence
of one gluon exchange potential. The variable $\lambda$ can be
interpreted as size of the quark. In Ref.~\cite{DP} the pointlike
quark-gluon vertex is replaced by a form factor,
$1/(q^{2}+\beta^{2})$ in which $\beta^{-1}$ is the effective quark
size to eliminate the singularity. In this work we have used both
regularized form and Gaussian form factor for coulomb and
$f_{c}\,\alpha_{s}\textbf{p}^2/(m^2r)$ parts of the potential which
cause the convergence of numerical results faster. Therefore, the
final form of the potential in the momentum-helicity space is
written as:
\begin{eqnarray}\label{eq8}
V^{S}_{\mathrm{\Lambda}\mathrm{\Lambda}'}(\textbf{p},\textbf{p}')&\equiv&\langle\textbf{p}S\mathrm{\Lambda}|V|\textbf{p}'S\mathrm{\Lambda}'\rangle=\sigma\,\langle\hat{\textbf{p}}S\mathrm{\Lambda}|\hat{\textbf{p}}'S\mathrm{\Lambda}'\rangle\Big[\delta(\textbf{q})\,r_{c}+\frac{1}{2\pi^{2}q^{4}}\big(2\cos(q\,r_{c})-2+q\,r_{c}\sin(q\,r_{c})\big)\Big]\nonumber\\
&&+\,f_{c}\alpha_{s}\,e^{-\lambda^{2}
q^{2}}\,\langle\hat{\textbf{p}}S\mathrm{\Lambda}|\hat{\textbf{p}}'S\mathrm{\Lambda}'\rangle\Biggl\{\frac{1}{r_{c}}\,\delta(\textbf{q})+\frac{1}{2\pi^{2}q^{2}}\Big(1-\frac{\sin(q\,r_{c})}{q\,r_{c}}\Big)\big(1+\frac{p^2}{m^{2}}\big)\nonumber\\&&-\frac{1}{8\pi^{2}m^{2}}-\frac{1}{12\pi^{2}m^{2}}\big(2S(S+1)-3\big)\nonumber\\[2mm]&&+\,\frac{3}{8\pi^{2}m^{2}}\frac{p\,p'}{q^{2}}\Big[\gamma
S(S+1)-2\mathrm{\Lambda}\mathrm{\Lambda}'-\frac{1}{\gamma}\big(S(S+1)-2\mathrm{\Lambda}'^{2}-2\mathrm{\Lambda}^{2}-2\mathrm{\Lambda}'^{2}\mathrm{\Lambda}^{2}\big)\Big]\nonumber\\
&&-\,\frac{1}{24\pi^{2}m^{2}q^{2}}\Big[6p\,p'\mathrm{\Lambda}\mathrm{\Lambda}'+2p'^{2}\big(S(S+1)-3\mathrm{\Lambda}'^{2}\big)+2p^{2}\big(S(S+1)-3\mathrm{\Lambda}^{2}\big)\nonumber\\&&-p\,p'\gamma
S(S+1)-\,3\frac{p\,p'}{\gamma}\big(S(S+1)-2\mathrm{\Lambda}'^{2}-2\mathrm{\Lambda}^{2}-2\mathrm{\Lambda}'^{2}\mathrm{\Lambda}^{2}\big)\Big]\Biggl\},
\end{eqnarray}
where
$\gamma=\hat{\textbf{p}}'\cdot\hat{\textbf{p}}=\cos\theta\cos\theta'+\sin\theta\sin\theta'\cos(\varphi-\varphi')$
and $|\textbf{p};\hat{\textbf{p}}S\mathrm{\Lambda}\rangle$ is the
momentum-helicity basis state which is eigenstate of the helicity
operator $\textbf{S}\cdot\hat{\textbf{p}}$ as:
\begin{eqnarray}\label{eq2}
\textbf{S}\cdot\hat{\textbf{p}}|\textbf{p};\hat{\textbf{p}}S\mathrm{\Lambda}\rangle=\mathrm{\Lambda}|\textbf{p};\hat{\textbf{p}}S\mathrm{\Lambda}\rangle.
\end{eqnarray}
Also we have~\cite{h1}:
\begin{eqnarray}\label{eq8}
\langle\hat{\textbf{p}}S\mathrm{\Lambda}|\hat{\textbf{p}}'S\mathrm{\Lambda}'\rangle=\sum_{N=-S}^{S}\,e^{iN(\varphi-\varphi')}d_{N\mathrm{\Lambda}}^{S}(\theta)\,d_{N\mathrm{\Lambda}'}^{S}(\theta').
\end{eqnarray}
If the vector \textbf{p} is along $z$-direction, it is clear that
the Eq.~(15) is reduced to:
\begin{eqnarray}\label{eq8}
\langle
\hat{\textbf{z}}S\Lambda|\hat{\textbf{p}}'S\Lambda'\rangle=\,e^{-i\Lambda\varphi'}\,d_{\Lambda\Lambda'}^{S}(\theta').
\end{eqnarray}
For numerical calculations we need the matrix elements of the
potential $V^{S}_{\Lambda\Lambda'}(p,p',\theta')$. These matrix
elements is related to the matrix elements Eq.~(13) as follows:
\begin{eqnarray}\label{eq9}
V^{S}_{\Lambda\Lambda'}(p,p',\theta')=e^{i\Lambda\varphi'}\,\langle
p\,\textbf{z};\hat{\textbf{z}}S\Lambda|V|\textbf{p}';\hat{\textbf{p}}'S\Lambda'\rangle.
\end{eqnarray}
By considering Eqs.~(13), (16) and (17), the final form of the
matrix elements of the potential which is inserted in the numerical
calculations is written as:
\begin{eqnarray}\label{eq8}
V^{S}_{\Lambda\Lambda'}(p,p',\theta')&=&\sigma\,d_{\Lambda\Lambda'}^{S}(\theta')\Big[\delta(\textbf{q})\,r_{c}+\frac{1}{2\pi^{2}q^{4}}\big(2\cos(q\,r_{c})-2+q\,r_{c}\sin(q\,r_{c})\big)\Big]\nonumber\\
&&+\,f_{c}\alpha_{s}\,e^{-\lambda^{2}
q^{2}}\,d_{\Lambda\Lambda'}^{S}(\theta')\Biggl\{\frac{1}{r_{c}}\,\delta(\textbf{q})+\frac{1}{2\pi^{2}q^{2}}\Big(1-\frac{\sin(q\,r_{c})}{q\,r_{c}}\Big)\big(1+\frac{p^2}{m^{2}}\big)\nonumber\\&&-\frac{1}{8\pi^{2}m^{2}}-\frac{1}{12\pi^{2}m^{2}}\big(2S(S+1)-3\big)\nonumber\\[2mm]&&+\,\frac{3}{8\pi^{2}m^{2}}\frac{p\,p'}{q^{2}}\Big[\gamma
S(S+1)-2\mathrm{\Lambda}\mathrm{\Lambda}'-\frac{1}{\gamma}\big(S(S+1)-2\mathrm{\Lambda}'^{2}-2\mathrm{\Lambda}^{2}-2\mathrm{\Lambda}'^{2}\mathrm{\Lambda}^{2}\big)\Big]\nonumber\\
&&-\,\frac{1}{24\pi^{2}m^{2}q^{2}}\Big[6p\,p'\mathrm{\Lambda}\mathrm{\Lambda}'+2p'^{2}\big(S(S+1)-3\mathrm{\Lambda}'^{2}\big)+2p^{2}\big(S(S+1)-3\mathrm{\Lambda}^{2}\big)\nonumber\\&&-p\,p'\gamma
S(S+1)-\,3\frac{p\,p'}{\gamma}\big(S(S+1)-2\mathrm{\Lambda}'^{2}-2\mathrm{\Lambda}^{2}-2\mathrm{\Lambda}'^{2}\mathrm{\Lambda}^{2}\big)\Big]\Biggl\},
\end{eqnarray}
with $\gamma=\hat{\textbf{p}}'\cdot\hat{\textbf{z}}=\cos\theta'$.

\section{Discussion and numerical results}

For numerical calculations as a first step we have used the Gaussian
quadrature grid points to discretize the momentum and the angle
variables. The integration interval for the momentum is covered by
two different hyperbolic and linear mappings of the Gauss-Legendre
points from the interval [-1,+1] to the intervals $[0,
p_{2}]\bigcup\,[p_{2}, p_{max}]$ respectively as:
\begin{equation}\label{eq16}
p=\frac{1+x}{\frac{1}{p_{1}}+(\frac{2}{p_{2}}-\frac{1}{p_{1}})\,x},\,\quad\quad
p=\frac{p_{max}-p_{2}}{2}\,x+\frac{p_{max}+p_{2}}{2}.
\end{equation}
Then we have calculated the matrix elements of the potential
$V_{\Lambda\Lambda'}(p,p',\theta')$, from Eq.~(18). According to the
Eq.~(3) integration over the spherical angle variable $\theta'$, has
been done independently. Finally, we have solved the integral
equations~(4)-(8) as eigenvalue equations. The integration over
momentum variable is cut off at $q_{max}$ = 10 GeV. This selection
is carried out so that the numerical results do not depend on this
choice. The typical values for $p_{1}$ and $p_{2}$ are 1 GeV and 3
GeV, respectively. These selections are done till the total number
of grid points for momentum intervals are decreased. Other
selections can be done but by different grid points for momentum
variables.

The parameters of the potential model which are shown in Table~I are
fixed by a fit to the masses of the states $\eta_{c}$, $J/\psi$ and
$h_{c}$, similar to what is done in Ref.~9. The results of
charmonium mass spectrum are shown in Table~II. They are compared
with the experimental data and another theoretical work. From Eqs.
(7) and (8) it is clear that existence of the tensor term in the
potential mix $S$- and $D$- partial waves but this mixed as it is
shown in Table III is so weak. I show the mixed charmonium states in
Table~II by their dominant partial wave.

As a test of our numerical calculations we have shown convergence of
the results as a function of number of grid points
$\mathrm{N_{P1}}$, $\mathrm{N_{P2}}$ and $\mathrm{N_{\theta}}$ for
the momentum and angle variables in Table~IV. $\mathrm{N_{P1}}$,
$\mathrm{N_{P2}}$ are the number of grid points for the intervals
[0, $p_{2}$] and [$p_{2}$, $p_{max}$] respectively.
$\mathrm{N_{\theta}}$ is corresponding to number of grid points for
spherical angle variable. In our calculations we have chosen
$\mathrm{N_{P1}}$=100, $\mathrm{N_{P2}}$=100
$\mathrm{N_{\theta}}$=200 grid points for to achieve an acceptable
accuracy.

\section{SUMMARY and outlook}
In this paper we have extended an approach based on
momentum-helicity basis states for calculation of mass spectrum of
heavy mesons by solving nonrelativistic form of the
Lippmann-Schwinger equation. As an application we have used this
approach to obtain the mass spectrum of charmonium. The advantage of
working with helicity states is that states are the eigenstates of
the helicity operator appearing in the quark-aintiquark potential.
Thus, using the helicity representation is less complicated than
using the spin representation with a fixed quantization axis for
representation of spin dependent potentials. This work is the first
step toward for studying single, double, and triple heavy-flavor
baryons in the framework of the nonrelativistic quark model by
formulation of the Faddeev equation in the 3D momentum-helicity
representation. Furthermore, we can apply this formalism
straightforwardly for investigation of heavy pentaquark systems,
which can be considered as two-body (heavy meson, baryon) systems
with meson-nucleon potentials which is underway.
%
\begin{table}
\caption{Parameters of the model.}
\label{tab:1}       
\begin{tabular}{ll}
\hline\hline\noalign{\smallskip}
$\sigma~~[\mathrm{GeV/fm}]$ & ~~~~~~~~~~~~~~~~~~~1.222 \\
$\lambda~~[\mathrm{GeV^{-1}}]$ & ~~~~~~~~~~~~~~~~~~~0.3154 \\
$m~[\mathrm{GeV}]$ & ~~~~~~~~~~~~~~~~~~~1.269  \\
$\alpha_{s}$ & ~~~~~~~~~~~~~~~~~~~0.2863 \\
$r_{c}~[\mathrm{fm}]$ & ~~~~~~~~~~~~~~~~~~~10\\
\noalign{\smallskip}\hline\hline
\end{tabular}
\end{table}

%
\begin{table}
\caption{Comparison of the obtained charmonium mass spectrum  with
the experimental data and another work.}
\label{tab:1}       
\begin{ruledtabular}
\begin{tabular}{lllll}
$n^{2S+1}L_J$ & Candidate & Exp.~\cite{Ei} &~~ Ref.~\cite{jo}&~~~ Mass\,[MeV]\\
\noalign{\smallskip}\hline\noalign{\smallskip}
$1^1S_0$ & ~~~~~$\eta_{c}$ & $2980.4\pm1.2$&~~~2980  &~~~ 2980.4\\
$1^3S_1$ & ~~~~$J/\psi$ & $3096.916\pm0.011$&~~~3097  &~~~~3096.9\\
$1^1P_1$ & ~~~~~$h_{c}$ & $3526.21\pm0.25$&~~~3527  &~~~~3526.2\\
$1^3P_0$ & ~~~~~$\chi_{c0}$ & $3415.16\pm0.35$&~~~3430  &~~~~3397.4\\
$1^3P_1$ & ~~~~~$\chi_{c1}$ & $3510.59\pm0.10$&~~~3503  &~~~~3503.5\\
$2^1S_0$ & ~~~~~$\eta'_{c}$ & $3638\pm5$&~~~3674  &~~~~3683.1\\
$2^3S_1$ & ~~~~~$\psi'$ &$3686.093\pm0.034$ &~~ 3765&~~~~3760.8\\
$1^3D_1$ & ~~~~~$\psi''$  &$3770\pm2.4$ &~~  3855&~~~~3850.6\\
$3^3S_1$ & ~~~~~$\psi'''$ &$4040\pm10$&~~ 4291&~~~~4285.4\\
\end{tabular}
\end{ruledtabular}
\end{table}

%
\begin{table}
\caption{Percent of each partial wave in mixed charmonium states.}
\label{tab:1}       
\begin{ruledtabular}
\begin{tabular}{llll}
$n^{2S+1}L_J$ &$c\bar{c}$ &$ P_S\%~$ &$P_D\%~$\\
\noalign{\smallskip}\hline\noalign{\smallskip}
$1^3S_1~(1^3S_1-1^3D_1)$ & $J/\psi$ & $99.93$&0.07  \\
$2^3S_1~(2^3S_1-2^3D_1)$ & $\psi'$ &$99.90 $ & 0.10\\
$3^3S_1~(3^3S_1-3^3D_1)$ & $\psi'''$ &$99.88$& 0.12\\
$1^3D_1\,(1^3D_1-1^3S_1)$ & $\psi''$  &$99.88$ &  0.12\\
\end{tabular}
\end{ruledtabular}
\end{table}

%
\begin{table}
\caption{The calculated charmonium mass spectrum as function of the
number of grid points $N_{P1}$, $N_{P1}$ and $N_{\theta}$.}
\label{tab:1}       
\begin{ruledtabular}
\begin{tabular}{lllllllllllll}
$\mathrm{N_{p1}}$ &$\mathrm{N_{p2}}$&$\mathrm{N_\theta}$ & $\eta_{c}$ & $J/\psi$& $h_{c}$ &$\chi_{c0}$&$\chi_{c1}$&$\eta'_{c}$&$\psi'$&$\psi''$&$\psi'''$\\
\noalign{\smallskip}\hline\noalign{\smallskip}
100&100&$140$ &  $2980.602$&3096.942  &3526.244&3397.517&3503.552&368.3357&3760.778&3850.573&4285.415\\
100&100&$160$ &  $2980.420$&3096.951 &3526.226&3397.449&3503.541&3683.153&3760.782&3850.563&4285.410\\
100&100&$180$ &  $2980.370$&3096.954  &3526.221&3397.434&3503.539&3683.095&3760.784&3850.561&4285.409\\
100&100&$200$ &  $2980.356$&3096.954  &3526.219&3397.431&3503.538&3683.080&3760.784&3850.560&4285.409\\
100&100&$220$ & $2980.353$ & 3096.954&3526.219&3397.430&3503.538&3683.075&3760.784&3850.560&4285.409\\
100&100&$240$ & $2980.352$ &  3.096954&3526.219&3397.430&3503.538&3683.074&3760.784&3850.560&4285.409\\
100&100&$260$ & $2980.352$& 3.096954&3526.219&3397.430&3503.538&3683.074&3760.784&3850.560&4285.409\\
\noalign{\smallskip}\hline
100&60&$200$ &  $2975.716$&3097.131 &3525.746&3397.449&3394.900&3503.182&3677.378&3850.232&4284.965\\
100&80&$200$ &  $2980.218$&3096.961 &3526.205&3397.376&3503.530&3682.920&3760.787&3850.553&4285.405\\
100&100&$200$ &  $2980.356$&3096.954 &3526.219&3397.431&3503.541&3683.538&3760.080&3850.560&4285.409\\
100&120&$200$ &  $2980.356$&3096.954 &3526.219&3397.431&3503.541&3503.538&3683.080&3850.560&4285.409\\
\noalign{\smallskip}\hline
50&100&$200$ &  $2980.356$&3096.954 &3526.219&3397.429&3503.538&3683.080&3760.784&3850.558&4285.409\\
80&100&$200$ &  $2980.356$&3096.954 &3526.219&3397.430&3503.538&3683.080&3760.784&3850.560&4285.409\\
100&100&$200$ &  $2980.356$&3096.954 &3526.219&3397.431&3503.541&3683.080&3760.784&3850.560&4285.409\\
\end{tabular}
\end{ruledtabular}
\end{table}

\section*
{Conflict of Interests}
The author declares that there is no conflict of interests regarding
the publication of this paper.

\appendix

\section{Fourier transformation of the regularized linear confining and Coulomb parts of the
potential} The three-dimensional Fourier transformation of the
potential $V (r)$ is defined as:
\begin{eqnarray}\label{eq9}
V(\textbf{p},\textbf{p}')=\frac{1}{2\pi^2q^2}\int_0^\infty dr\, r
V(r) \sin qr,
\end{eqnarray}
where $\textbf{q}=|\textbf{p}-\textbf{p}'|$. Fourier transformation
of the regularized linear confining and Coulomb parts of the
quark-antiquark potential is written as:
\begin{eqnarray}\label{eq9}
V(\textbf{p},\textbf{p}')&=&\frac{1}{2\pi^2q^2}\Bigg\{\int_0^{r_{c}}
dr\, r \,V(r) \sin qr+V(r_c)\int_{r_{c}}^{\infty} dr\, r  \sin
qr\Bigg\} \nonumber \\&=&\frac{1}{2\pi^2q^2}\Bigg\{\int_0^{r_{c}}
dr\, r \,V(r) \sin qr+V(r_c)\int_{0}^{\infty}  dr \,r\,\sin
qr-V(r_c)\int_{0}^{r_{c}} dr\, r  \sin qr\Bigg\}\nonumber
\\&=&\frac{1}{2\pi^2q^2}\Bigg\{\int_0^{r_{c}} dr\, r \,V(r) \sin
qr+V(r_c)\,\delta(\textbf{q})-V(r_c)\int_{0}^{r_{c}} dr\, r  \sin
qr\Bigg\},
\end{eqnarray}
where potential is kept fixed at cutoff $r_c$. Therefore inserting
the linear $V(r)=\sigma r$, and Coulomb $V(r)=f_{c}\,\alpha_{s}/r$,
parts of quark-antiquark potential in above equation and calculation
of corresponding integrals analytically, yields:
\begin{eqnarray}\label{eq9}
V(\textbf{p},\textbf{p}')&=&\sigma
\,\Big[\,\delta(\textbf{q})\,r_{c}+\frac{1}{2\pi^{2}q^{4}}\big(2\cos(q\,r_{c})-2+q\,r_{c}\sin(q\,r_{c})\big)\Big],
\\ \nonumber\\ V(\textbf{p},\textbf{p}')
&=&f_{c}\,\alpha_{s}\Big[\frac{\delta(\textbf{q})}{r_{c}}+\frac{1}{2\pi^{2}q^{2}}\big(1-\frac{\sin(q\,r_{c})}{q\,r_{c}}\big)\Big].
\end{eqnarray}


\begin{thebibliography}{99}

\bibitem{h1} I. Fachruddin, Ch. Elster, and W. Gl\"{o}ckle, Phys. Rev. C \textbf{62}, 044002 (2000).

\bibitem{h2} I. Fachruddin, Ch. Elster, and W. Gl\"{o}ckle, Phys. Rev. C \textbf{63}, 054003 (2001).

\bibitem{RN}  M. Radin and N. Tazimi, Phys. Rev. D \textbf{90}, 085020 (2014).

\bibitem{Sur} Suraj N. Gupta, Stanley F. Radford, and Wayne W. Repko
 Repko, Phys. Rev. \textbf{D} 26 (1982), 3305.

\bibitem{ML} M. R. Hadizadeh and Lauro Tomio, AIP Conf. Proc. 1296, 334 (2010).

\bibitem{J} J. Carlson, J. B. Kogut, and V. R. Pandharipande, Phys. Rev. D \textbf{28}, 2807 (1983).

\bibitem{DP} D. P. Stanley and D. Robson, Phys. Rev. \textbf{D} 21, 3180 (1980).

\bibitem{Ei} S. Eidelman et al., Physics Letters B \textbf{592} 1 (2004).

\bibitem{jo} J. Eiglsperger, \emph{Quarkonium Spectroscopy: Beyond One-Gluon
Exchange}, Diploma thesis, Technische Universit\"{a}t M\"{u}nchen,
2007, arXiv:0707.1269 [hep-ph]


\end{thebibliography}
\end{document}